\documentclass[12pt,letterpaper]{article}
\usepackage{epsfig,pslatex}
\begin{document}

\vspace{4cm}
\begin{center}
{\Large\bf{Probing the Anomalous Couplings of the Top Quark with Gluon at the LHC and Tevatron}}\\

\vspace{1cm}
 {\bf Hoda Hesari and Mojtaba Mohammadi Najafabadi }  \\
\vspace{0.5cm}
{\sl  School of Particles and Accelerators, \\
Institute for Research in Fundamental Sciences (IPM) \\
P.O. Box 19395-5531, Tehran, Iran}\\

\vspace{2cm}
 \textbf{Abstract}\\
 \end{center}

In this paper, we study the sensitivity of the fraction of $t\bar{t}$
events arising from gluon-gluon fusion to the
chromoelectric and chromomagnetic dipole moments (CEDM and CMDM) as
well as the total and differential $t\bar{t}$ cross sections at the LHC and
Tevatron. The sensitivity of measured charged asymmetry at the LHC to
CEDM and CMDM is also studied.
We find that at the Tevatron and the LHC, non-zero values of CMDM could suppress
the $t\bar{t}$ production rate.
It is shown that the ratio of $\sigma(gg\rightarrow
t\bar{t})/\sigma(p\bar{p}\rightarrow t\bar{t})$ at the Tevatron
is more sensitive to CEDM and CMDM than the LHC case.
The presence of CEDM always increases the contribution of
gluon-gluon fusion process in top pair rate at the Tevatron and
LHC. Except for a small range of CMDM, the presence of CEDM and CMDM
can increase the fraction of gluon-gluon fusion at the Tevatron and LHC.
The measured ratio of $\sigma(gg\rightarrow t\bar{t})/\sigma(p\bar{p}\rightarrow t\bar{t})$ at the Tevatron
is used to derive bounds on the chromoelectric and chromomagnetic
dipole moments as well as the total and differential ($d\sigma/dm_{t\bar{t}}$) cross sections
at the LHC and Tevatron, and the measured charged asymmetry at the
LHC. The combination of $d\sigma_{TeV}/dm_{t\bar{t}}$ and
$\sigma_{LHC}$ provides stringent limits on CMDM and CEDM.
\vspace{1cm}\\
PACS number(s): 14.65.Ha, 12.60.-i

\newpage

\section{Introduction}

From the point of view of the large mass of the top quark, it could provide excellent
probes of the electroweak symmetry breaking mechanism as well as
clarifying the nature of any force and any particle responsible for this
phenomenon. Furhtermore, the top quark production and decay rates can
also prepare a good place to probe possible new parity violating and
anomaluos CP violating interactions which could be induced by non-SM
particles \cite{wagner},\cite{werner},\cite{beneke}.

The top quark couplings still need to be measured
more precisely to be able to observe any deviations from the standard model
predictions. In particular, the anomalous interactions between the top
quark and gauge bosons can influence the top quark production cross sections
at high energy colliders such as the Large Hadron Collider (LHC) and Tevatron.
The beyond standard model effect in the production of $t\bar{t}$ could be parameterized in a rather model independent way
by effective couplings of quark-gluon, which indeed  may be induced by
exchange of heavy particles.
The $t\bar{t}$ production at hadron colliders is obviously
the only place for {\it direct} search of anomalous $gt\bar{t}$ interactions. In this analysis, in particular, we concentrate on the study of
top quark chromomagnetic and chromoelectric couplings that are described by the following effective Lagrangian:
\begin{eqnarray}\label{lag}
\mathcal{L}_{t\bar{t}g} = \frac{ig_{s}}{2m_{t}}\bar{t}\sigma_{\mu\nu}q^{\nu}(\kappa-i\tilde{\kappa}\gamma_{5})tG^{\mu}
\end{eqnarray}
where $g_{s}$ is the strong coupling constant, $q^{\nu}$ is the 4-momentum of the involved gauge boson (gluon), and $G^{\mu}$ is the
gluon gauge field.
It is worth mentioning that due to the excellent agreement between the SM expectations and
the present experimental data, any deviations from the SM are small.
Accordingly, the new effective terms must be very small and the interference term
between SM and new effective terms could be big enough to be measured.
Please note that $\kappa$ and $\tilde{\kappa}$ are loosely referred as the CMDM and CEDM form factors but they are related
to the dimensionful CMDM and CEDM via the following equations:
\begin{eqnarray}
\kappa = \frac{2m_{t}}{g_{s}}\mu^{g}_{t},~\tilde{\kappa} = \frac{2m_{t}}{g_{s}}d^{g}_{t}
\end{eqnarray}

In renormalizable theories $\kappa$ and $\tilde{\kappa}$ are induced at loop level.
Notice that non-zero value for $\tilde{\kappa}$ leads to CP violating interactions. The generated chromoelectric moment
(CEDM) by the CKM phase is very small.
Therefore, the measurement of observables that signify CP violation in top pair
production using large statistics
data should be studied carefully.
In the SM, at one-loop level the QCD corrections generate CMDM via gluon exchange in
two distinct Feynman diagrams. One of the diagrams is quite similar to the QED case (with replacement of
photon with gluon) and another one consists of an external gluon
which is coupled to the internal gluons which is because of the
non-abelian nature of QCD. Similar to QED, these diagrams generate CMDM
proportional to $\alpha_{s}/\pi$ (but after replacing $\alpha_{em}$ by $\alpha_{s}$).
There is an overall factor which is originating from the multiplication
of color matrices. It should be indicated that in addition to QCD corrections, Higgs and
$Z$ boson exchanges generate CMDM \cite{higgs1},\cite{higgs2}.

So far, there have been several studies on the anomalous chromoelectric
and chromomagnetic dipole moments. The CEDM and CMDM effects on the
top pair and single top cross sections and other related observables at hadron colliders were examined in \cite{k1},
\cite{k2},\cite{k3},\cite{k4},\cite{k5},\cite{k6},\cite{k61},\cite{k7}.
It was shown that the $t\bar{t}$ differential cross section (such as transverse momentum and
invariant mass of $t\bar{t}$)
is sensitive to the sign of CMDM as well as its size.
Possible range on $\kappa$ using 30 fb$^{-1}$ of future
$e^{-}e^{+}$ collider  with 500 GeV center-of-mass energy
is $-2.1<\kappa<0.6$ \cite{NLC}.
Apart from the direct study of CMDM and CEDM at hadron colliders in $t\bar{t}$ events,
they can modify the decay rate of $B\rightarrow X_{s}\gamma$ \cite{bsg1},\cite{bsg2} at loop level.
Using the measured branching ratio of $Br(b\rightarrow s\gamma)$, tight bounds on $\kappa$ was extracted. The
limit is $-0.03 < \kappa < 0.01$.
In specific models, such as minimal supersymmetric standard model (MSSM), little Higgs model,
and two-Higgs doublet model, non-zero $\kappa$ could be generated
\cite{mssm},\cite{2hdm},\cite{LH}.
For example, in a specific parameter space of MSSM, the corrections to
CMDM can be as large as $20\%$.
In \cite{LH}, the authors have calculated the one-loop contributions
of the new particles of the littlest Higgs model with T-parity to the top quark chromomagnetic dipole moment (CMDM).
It was shown that the CMDM that is generated by this model is one order of magnitude smaller than
the SM predicted value.
In \cite{unparticle}, the effect of unparticle to the chromomagnetic dipole moment (CMDM) of
the top quark has been studied. The induced  effects by scalar and vector unparticle operators were computed on the CMDM.
It was shown that depending on the parameters space and couplings,
unparticle could suppress the chromomagnetic dipole moment.
In \cite{k4}, a class of technicolor models was proposed that contains techniscalars
which may produce large values of chromomagnetic dipole moments for the top quark as well as
examining the dependency of the differential and the cross section of $t\bar{t}$ on $\kappa$.

In this paper, in addition to the total and differential cross sections of
$t\bar{t}$ at the LHC and Tevatron, we use the measured ratio
of cross section of $t\bar{t}$ arising from gluon-gluon fusion to the
total $t\bar{t}$ cross section to study
the allowed region of parameters $\kappa$ and $\tilde{\kappa}$.
The sensitivity of measured charge asymmetry in top pair events at the LHC
is also examined to the anomalous couplings of top quark to gluon.
Using the present differential cross section in invariant mass of
$t\bar{t}$ ($d\sigma/dm_{t\bar{t}}$) and the total LHC cross section,
stringent bounds on CMDM and CEDM are extracted.

\section{Influence of CEDM and CMDM on the $t\bar{t}$ Cross Section}

The effective Lagrangian approach is specially useful when the underlying new physics is not known.
The effective Lagrangian for describing the interaction between the top quark and a gluon, which considers
the CEDM and CMDM form factors of the top quark, was introduced in Eq.\ref{lag}. An important point is that
the effective interaction of $ttgg$ which is absent in the SM must be taken into account to ensure gauge
invariance.

The $t\bar{t}$ production at hadron colliders $(pp(\bar{p})\rightarrow t\bar{t})$ can proceed at partonic level through
quark-antiquark annihilation $(q\bar{q}\rightarrow t\bar{t})$ or gluon-gluon fusion $(gg\rightarrow t\bar{t})$.
The parton level cross sections for $gg\rightarrow t\bar{t}$ and $q\bar{q}\rightarrow t\bar{t}$
has the following forms \cite{k1}:
\begin{eqnarray}
  \frac{d\hat{\sigma}_{q\bar{q}\rightarrow t\bar{t}}}{d\hat{t}} &=&  \frac{8\pi\alpha_{s}^{2}}{9\hat{s}^{2}}
  [ \frac{1}{2}-f(\hat{s},\hat{t})+g(\hat{s})-\kappa+\frac{\kappa^{2}}{4}(1+\frac{f(\hat{s},\hat{t})}{g(\hat{s})})+
  \frac{\tilde{\kappa}^{2}}{4}(\frac{f(\hat{s},\hat{t})-1}{g(\hat{s})}) ] \\
   \frac{d\hat{\sigma}_{gg\rightarrow t\bar{t}}}{d\hat{t}}&=& \frac{\pi\alpha_{s}^{2}}{12\hat{s}^{2}} [(\frac{4}{f(\hat{s},\hat{t})}-9)(\frac{1}{2}-f(\hat{s},\hat{t})+2g(\hat{s})(1-\frac{g(\hat{s})}{f(\hat{s},\hat{t})})
   -\kappa(1-\frac{\kappa}{2})) \\ \nonumber
   &+&\frac{1}{4}(\kappa^{2}+\tilde{\kappa}^{2})
   (\frac{7}{g(\hat{s})}(1-\kappa)+\frac{1}{2f(\hat{s},\hat{t})}(1+\frac{5\kappa}{2})) \\ \nonumber
   &+&\frac{1}{16}(\kappa^{2}+\tilde{\kappa}^{2})^{2}(\frac{1}{f(\hat{s},\hat{t})}-\frac{1}{g(\hat{s})}+\frac{4f(\hat{s},\hat{t})}{g^{2}(\hat{s})})]
\end{eqnarray}
where
\begin{eqnarray}
f(\hat{s},\hat{t}) = \frac{(\hat{t}-m_{t}^{2})(\hat{u}-m_{t}^{2})}{\hat{s}^{2}},~g(\hat{s})=\frac{m_{t}^{2}}{\hat{s}^{2}},~\hat{s}+\hat{t}+\hat{u}=2m_{t}^{2}
\end{eqnarray}
According to the parton level cross section, in spite of CMDM $(\kappa)$, the cross-section is expected to be symmetric with respect to
CEDM $(\tilde{\kappa})$. Since the cross section is not a CP violating observable, $\tilde{\kappa}$ enters in the cross section in even powers.

It is notable that the contribution of the CEDM and CMDM to the partonic differential cross sections
of $gg\rightarrow t\bar{t}$ and $q\bar{q}\rightarrow t\bar{t}$ grows
with $\hat{s}$. In general, the new couplings could be dependent on
$\hat{s}$. However, if the  new physics scale is much higher than
$\sqrt{\hat{s}}$, we can neglect any dependence of CEDM and CMDM on
$\hat{s}$. The increase of center-of-mass energy reduces the
reliability of the assumption of constantness of the CEDM and CMDM.

The hadronic cross section can be obtained by convoluting the parton level cross section with parton distribution
functions:
\begin{eqnarray}
d\sigma(pp(\bar{p})\rightarrow t\bar{t}) = \sum_{ij=q\bar{q}(gg)}\int_{0}^{1}dx_{1}\int_{0}^{1}dx_{2}(f_{i}(x_{1},Q^{2})f_{j}(x_{2},Q^{2})+i\longleftrightarrow j)d\hat{\sigma}_{ij}
\end{eqnarray}
where $f_{i}(x,Q^{2})$ are the parton distribution functions (PDF's).
The parton distribution functions of CTEQ6L set \cite{cteq} with the $Q$-scale is equal to the top quark mass
are used to perform the calculations. The top quark mass has been taken to be 173 GeV.

\begin{figure}
\centering
  \includegraphics[width=7cm,height=5cm]{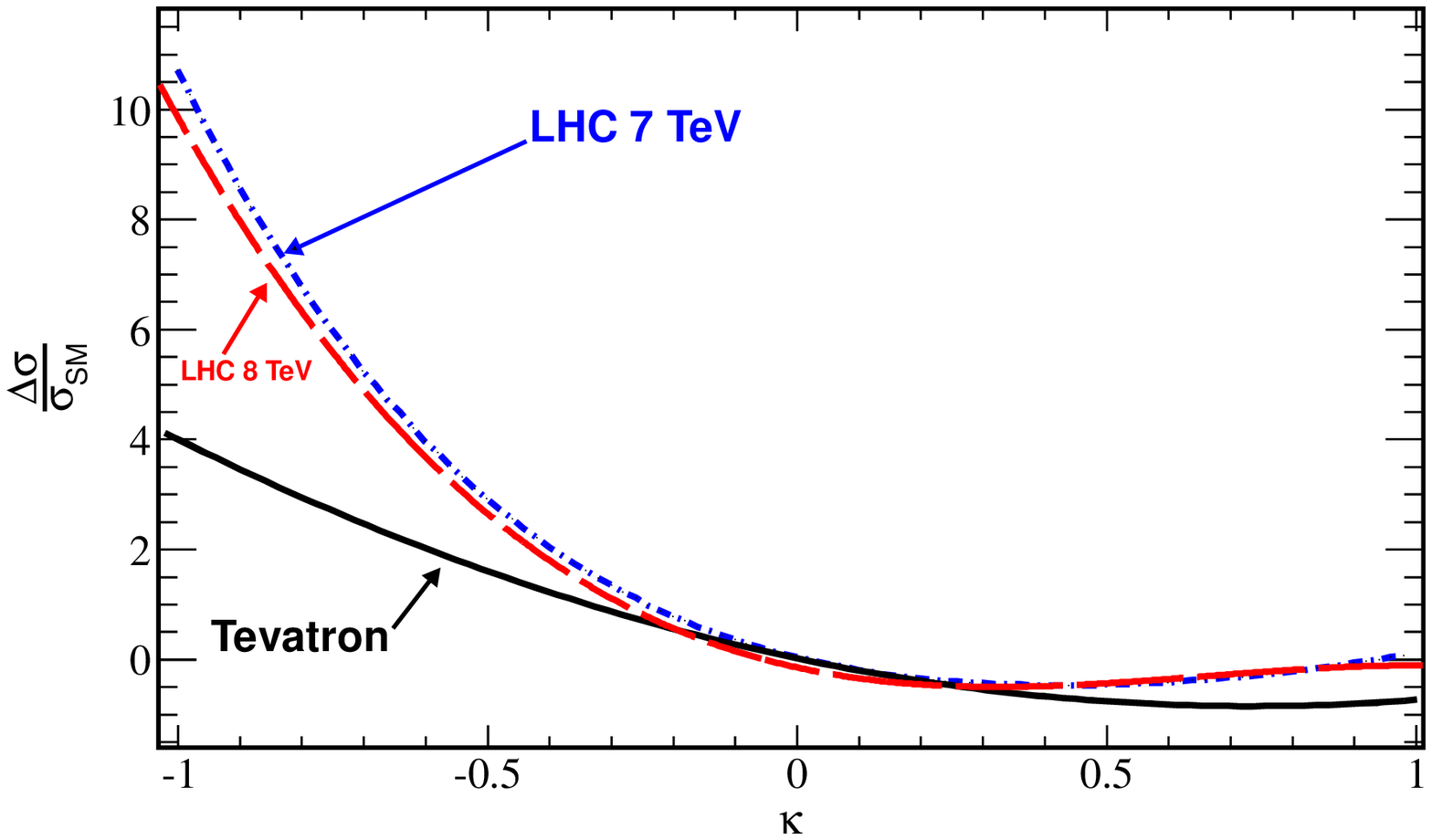}
  \includegraphics[width=7cm,height=5cm]{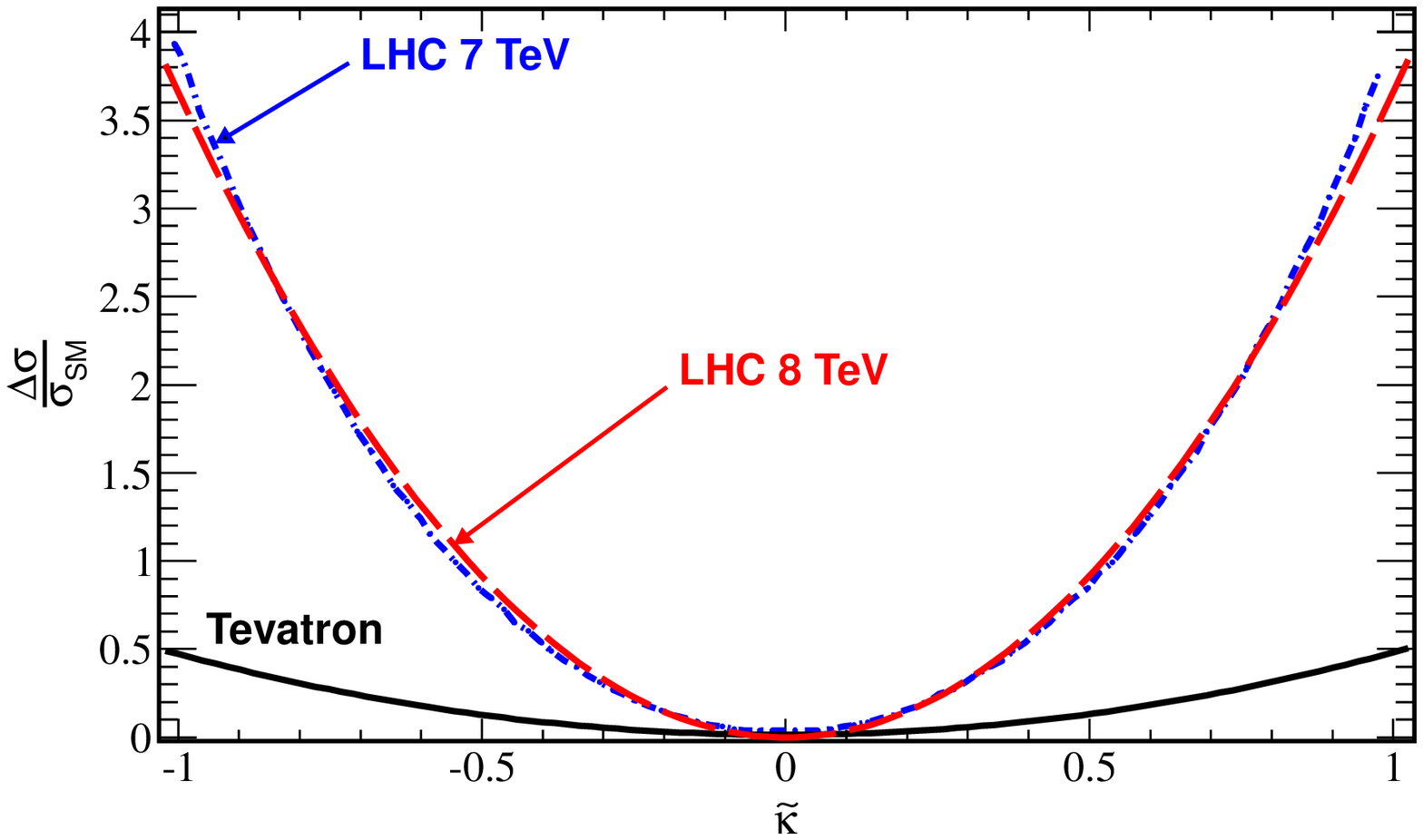}
  \caption{The relative change to the $t\bar{t}$ production cross section at the Tevatron and LHC7 in terms of $\kappa$ (left) and $\tilde{\kappa}$ (right).
}\label{sigma}
\end{figure}

Figures \ref{sigma} depict the dependence of the relative change of
the total $t\bar{t}$ cross section originating from CMDM and CEDM at
the Tevatron, LHC7, and LHC8.
As it can be seen, the sensitivity of the cross section at the LHC to
CMDM and CEDM is higher than the Tevatron rate.
The left plot in Fig. \ref{sigma} shows the dependence of the relative change of total cross section on $\kappa$ when $\tilde{\kappa} =0$.
For positive values of $\kappa$ in the range of around $0 < \kappa < 0.9 (1.6)$ for the LHC (Tevatron), the cross sections
are suppressed. The $t\bar{t}$ cross sections are decreased up to the level of $50\%$ when $0 < \kappa < 0.9 (1.6)$ for the LHC (Tevatron).
As discussed previously and can be seen in the right plot of Fig.\ref{sigma}, the cross section is symmetric
when $\tilde{\kappa}\rightarrow -\tilde{\kappa}$.

Any measured value of
the $t\bar{t}$ cross section significantly deviated from the SM
predictions could lead to the existence of a non-zero
value for CEDM and/or CMDM. The recent experimental measured values of the cross section of $t\bar{t}$ at the LHC at the center-of-mass
energy of 7 TeV (using 2.3 fb$^{-1}$ of data) and Tevatron (obtained
from 4.6 fb$^{-1}$ of data) including all sources of uncertainties are \cite{lhc},\cite{tevatron}:
\begin{eqnarray}
\sigma(pp\rightarrow t\bar{t})_{LHC} = 161.9 \pm 6.6 ~,~\sigma(p\bar{p}\rightarrow t\bar{t})_{Tevatron} = 7.5\pm 0.48
\end{eqnarray}
These measurements are in agreement with the standard model predictions which are $163$ pb for the LHC7 and
$7.08$ pb for the Tevatron \cite{sigmatheory}. The present measured
value for the $t\bar{t}$ cross section at the LHC at 8 TeV is
$228.4\pm 32$ pb. This measurement has still large
uncertainty \cite{lhc8}. Therefore, we do not include it in our analysis.
Using the current measurements and the relative uncertainties, the following constraints are extracted:
\begin{eqnarray}
LHC7:-0.03<\kappa<0.92~,-0.09<\tilde{\kappa}<0.09 \\ \nonumber
Tevatron:-0.03<\kappa<1.5~,-0.37<\tilde{\kappa}<0.37
\end{eqnarray}
In obtaining the bounds on each anomalous coupling, the other has been set to zero. As it can be seen, the
the cross section of LHC is more sensitive to the new interactions and provide tighter bounds with respect to
the Tevatron.
From fig.\ref{sigma}, we conclude that the corrections that the
$t\bar{t}$ cross section receives from $\kappa$ and $\tilde{\kappa}$
are almost similar in proton-proton collisions at 7 TeV and 8 TeV.
In \cite{k1}, it was shown that the shape of transverse momentum
and rapidity of top (anti-top) quark are not sensitive to
$\tilde{k}$. While the top quark transverse momentum is sensitive to
the sign of CMDM ($\kappa$).
In \cite{k7}, the effects of CMDM and CEDM on the energy,
transverse momentum, and angular distributions of lepton in top decay
have been investigated. Except for the size , no change in shape is observed.

\begin{figure}
\centering
  \includegraphics[width=7cm,height=5cm]{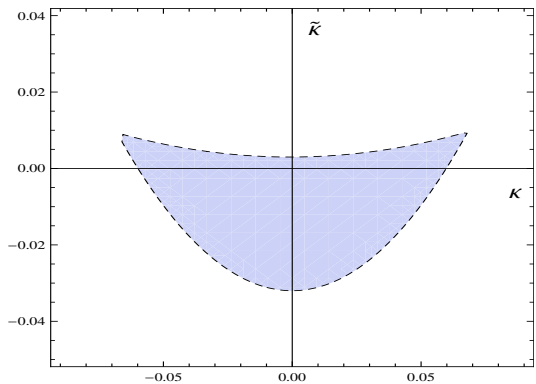}
  \caption{The $68\%$ C.L. region obtained from a $\chi^{2}$ fit to
    $d\sigma_{TeV}/dm_{t\bar{t}}$
 and the total cross section of top pair measured by CMS
    experiment. Horizontal axis denotes $\tilde{k}$ and vertical axis
    is for $\kappa$.
}\label{mttb}
\end{figure}

The $t\bar{t}$ differential cross section with the invariant mass of
$t\bar{t}$ has been also performed by the CDF experiment at the
Tevatron \cite{dmtt}. This study has been done with 2.7 fb$^{-1}$ of
data. The $t\bar{t}$ spectrum was found to be consistent with the SM
expectation. The bin by bin measured values are shown in Table
\ref{tab_mtt}. Now, to constrain the top quark CEDM and CMDM, we
combine all bins of $d\sigma/dm_{t\bar{t}}$ presented in Table
\ref{tab_mtt} and the total cross section $t\bar{t}$ by the CMS
experiment at the LHC into a global $\chi^{2}$ fit. The $68\%$
C.L. region in the $\kappa,\tilde{\kappa}$ plane is depicted in Fig.\ref{mttb}.
As it can be seen, we get stringent bounds on CMDM and CEDM: 
\begin{eqnarray}
-0.032<\kappa<0.01~,-0.063<\tilde{\kappa}<0.063
\end{eqnarray}
These bounds are compatible with the bounds obtained in \cite{k7} using
the total cross section of $t\bar{t}$ at the LHC and Tevatron. We have
obtained a bit tighter limits because of using more data information with
low uncertainties.

\begin{table}
\begin{center}
\begin{tabular}{|c|c|c|}
  \hline
    bin (GeV) & $\sigma_{TeV}$ (CDF) pb &  $\sigma_{TeV}$ (NLO-SM)
    pb\\ \hline
    350-400 & $ 3.115\pm 0.559 $  &  2.45\\
    400-450 & $ 1.690\pm 0.269 $  &  1.90\\
    450-500 & $ 0.790\pm 0.170 $  &  1.15\\
    500-550 & $ 0.495\pm 0.114 $  &  0.60\\
    550-600 & $ 0.285\pm 0.071 $  &  0.40\\
    600-700 & $ 0.239\pm 0.073 $  &  0.31\\
    700-800 & $ 0.080\pm 0.037 $  &  0.10\\
    800-1400 & $0.041\pm 0.021 $  &  0.036\\
  \hline
\end{tabular}
\caption{\label{tab_mtt} The CDF measurement of
  $d\sigma/dm_{t\bar{t}}$, the SM values are at NLO.}
\end{center}
\end{table}

\section{The Contribution of Gluon-Gluon Fusion in $t\bar{t}$ Production}

In spite of proton-proton collisions at the LHC,
in the Tevatron in proton-antiproton collisions
the main production mechanism of $t\bar{t}$ is quark-antiquark annihilation.
In particular, around $90\%$ of the $t\bar{t}$ cross section at the Tevatron is coming from the quark-antiquark annihilation.
The fractions of $t\bar{t}$ production from gluon-gluon fusion and
quark-antiquark annihilation are strongly dependent on the choice of
parton distribution functions \cite{cacciari}. Therefore, precise
measurement of these fractions provide better understanding of parton
distribution functions. Furthermore, there are several models beyond
the standard model which directly affect the $t\bar{t}$ production
mechanism. The measurement of these fractions allow us to probe those
models \cite{ggbsm1}.

The CDF Collaboration has measured the gluon-gluon contribution
to the top pair corss section at the Tevatron.
In order to measure the ratio $R=\sigma(gg\rightarrow
t\bar{t})/\sigma(p\bar{p}\rightarrow t\bar{t})$, some variables that are
sensitive to the $t\bar{t}$ production mechanism are used.
Since the functional form of the cross section is different for
gluon-gluon fusion from $q\bar{q}$ annihilation, some variables could
be found to distinguish between the production mechanisms.
For example, the cosine of the angle between the momentum of the top quark and the
incoming proton direction, the velocity of the top quark are of the
variables which are used to distinguish between the production mechanisms.
Since top pair events with parallel top-quark spins come exclusively
from gluon-gluon production, the angular distributions of the top
decay products show different behavior with respect to those coming
from $q\bar{q}$ annihilation. More details of the distinguishing
variables are described in \cite{gg}. In \cite{gg}, a measurement of
the ratio of the $t\bar{t}$ events produced through gluon-gluon fusion
to the total $t\bar{t}$ events is presented. Using around $1$
fb$^{-1}$ of data collected with the CDF detector and taking into
account only semi-leptonic $t\bar{t}$ events lead to:
\begin{eqnarray}
R = \frac{\sigma(gg\rightarrow t\bar{t})}{\sigma(p\bar{p}\rightarrow
  t\bar{t})} = 0.07^{+0.15}_{-0.07}.
\end{eqnarray}

\begin{figure}
\centering
  \includegraphics[width=7cm,height=5cm]{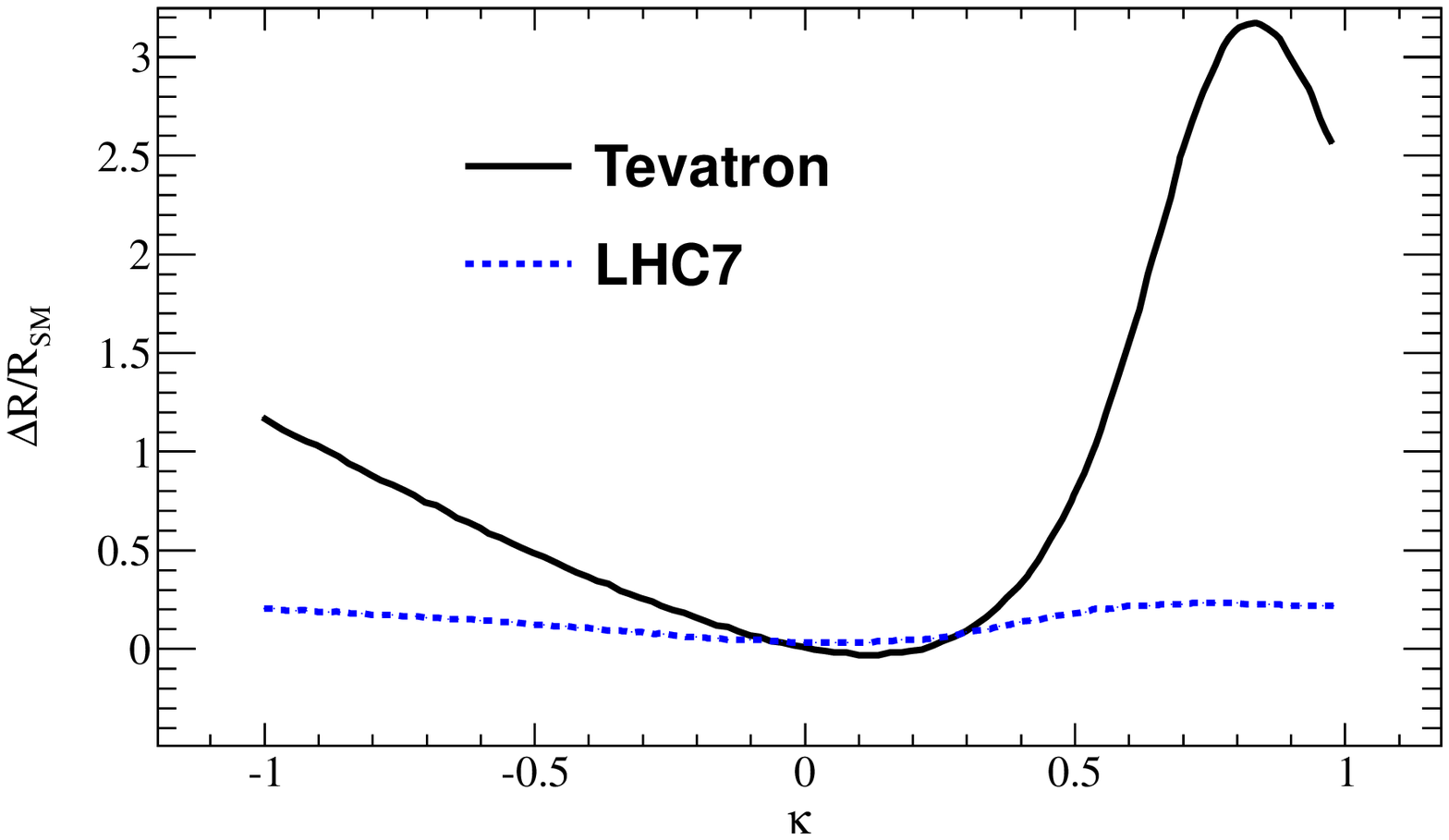}
  \includegraphics[width=7cm,height=5cm]{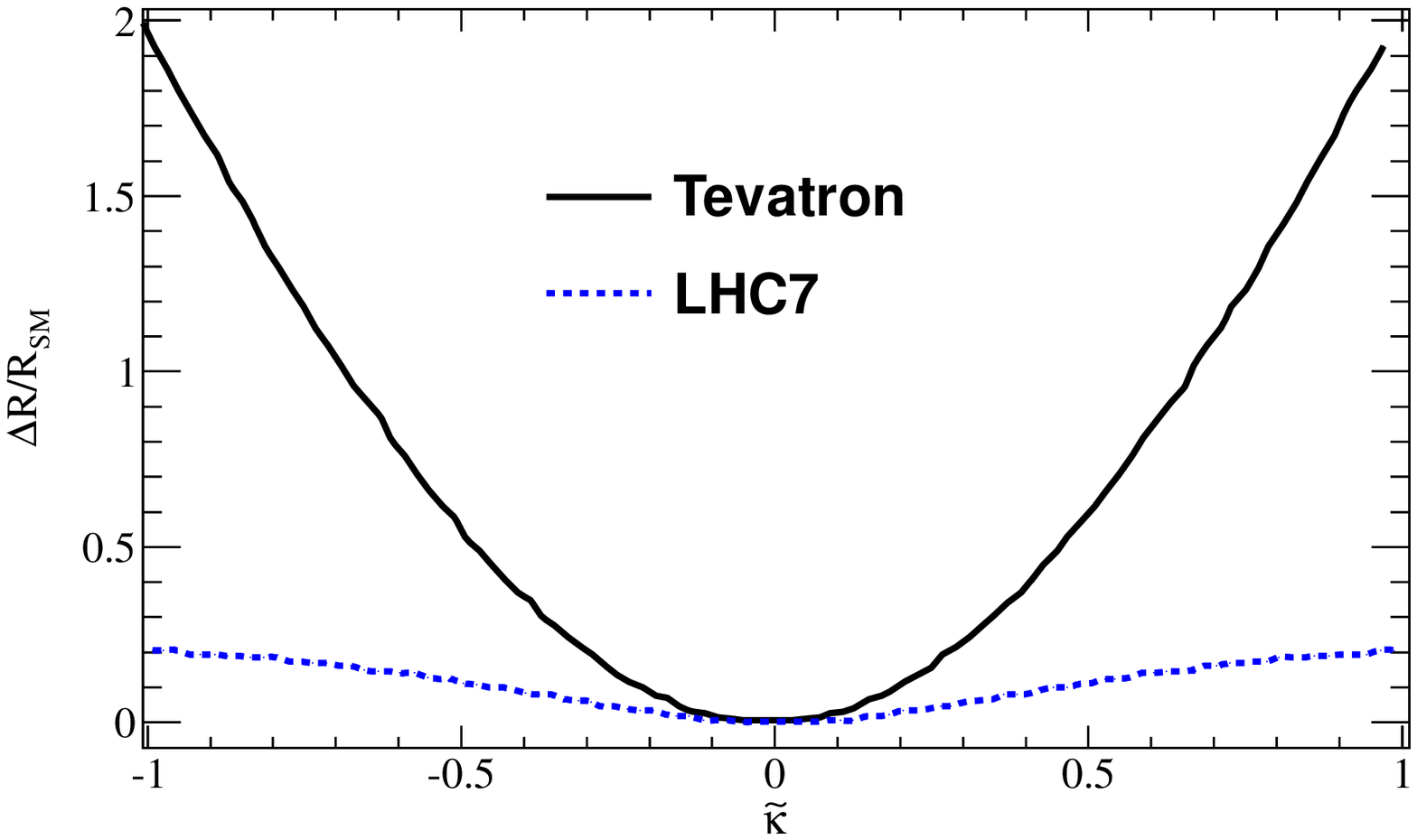}
  \caption{The relative correction originating from CMDM and CEDM to the ratio of $\sigma(gg\rightarrow t\bar{t})/\sigma(p\bar{p}\rightarrow t\bar{t})$.
}\label{ratio}
\end{figure}

Figs.\ref{ratio} depict the relative correction due to non-zero values
of CEDM and CMDM to the fraction of gluon-gluon fusion  in $t\bar{t}$
production at the LHC7 and Tevatron.
As it can be seen, $\Delta R = (R(\kappa,\tilde{\kappa})-R_{SM})/R_{SM}$
is significantly sensitive to $\kappa$ and $\tilde{\kappa}$ at the
Tevatron. While at the LHC, the presence of CEDM and CMDM
does not cause to considerable change in the contribution of
gluon-gluon fusion in top pair production.
The present Tevatron measurement of the ratio $R$ gives the following
limits:
\begin{eqnarray}
-1.1<\kappa<0.6~,~-0.8<\tilde{\kappa}<0.8 \nonumber
\end{eqnarray}
The current Tevatron measurement is based on around 1 fb$^{-1}$ of
data. However, Tevatron has taken more than 5 fb$^{-1}$ of data before
shut down. The analysis of full data improves the
measurement of $R$, for example a future measurement of the SM expectation for
$R$ with $10\%$ uncertainty gives:
\begin{eqnarray}
-0.15<\kappa<0.3~,~-0.18<\tilde{\kappa}<0.18 \nonumber
\end{eqnarray}
So far, there is no measurement of $R$ at the LHC. Assuming the SM
prediction with $10\%$ uncertainty leads to:
\begin{eqnarray}
-0.3<\kappa<0.25~,~-0.45<\tilde{\kappa}<0.45 \nonumber
\end{eqnarray}
%In fig.\ref{2dim}, we present the allowed region in the
%$(\kappa,\tilde{\kappa})$
%plane which satisfies the LHC cross section measurement and the
%measured $R$ by the CDF detector at the Tevatron.
%The region allowed by cross sections measured by CMS, ATLAS, CDF,
%and D0 collaborations altogether has been presented in \cite{k7}.
%The combination of all of these experiments leads to much tighter
%limits in the $(\kappa,\tilde{\kappa})$ plane.

%\begin{figure}
%\centering
%  \includegraphics[width=8cm,height=6cm]{kkt.eps}
%  \caption{The allowed region in the plane of $\kappa,\tilde{\kappa}$ which satifies the LHC cross section
%    measurement and the measured ratio $R=\sigma(gg\rightarrow
%    t\bar{t})/\sigma(p\bar{p}\rightarrow t\bar{t})$.
%}\label{2dim}
%\end{figure}

\section{Charge Asymmetry}

\begin{figure}
\centering
  \includegraphics[width=7cm,height=5cm]{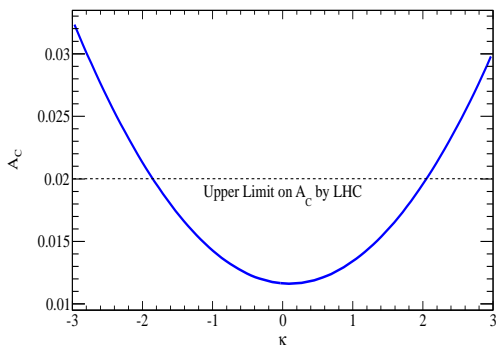}
  \caption{Charge asymmetry as a function of $\kappa$ and the upper
    limit on charge asymmetry measured by the CMS experiment.
}\label{ac}
\end{figure}

Since the initial state of the collisions at the LHC is proton-proton
which is symmetric; it is expected that the
rapidity distributions of top quarks and top antiquarks are
symmetrical around $y = 0$.
However, at the LHC the initial state quarks are abundantly valence
quarks, while always the antiquarks are
sea quarks. The larger average momentum fraction of the quarks leads
to an excess in production of top
quarks in the forward directions. This fact causes to a broader rapidity distribution of top quarks in the SM
with respect to the produced top antiquarks and therefore a charge
asymmetry is produced \cite{AC11}.
One way to define the top charge asymmetry is as the following:
\begin{eqnarray}
A_{C} = \frac{N(|y_{t}|>|y_{\bar{t}}|)-N(|y_{t}|<|y_{\bar{t}}|)}{N(|y_{t}|>|y_{\bar{t}}|)+N(|y_{t}|<|y_{\bar{t}}|)}
\end{eqnarray}
The NLO prediction for the charge asymmetry at a centre-of-mass energy of 7 TeV is
$A_{C}(theory) = 0.0115 \pm 0.0006$ \cite{AC11}. The existence of new sources of physics with different vector and axial-vector
couplings to top quarks and antiquarks could enhance these asymmetries
\cite{AC11}. The recent measured value by the CMS Collaboration using
$4.7$ fb$^{-1}$ of data is $A_{C} = 0.004 \pm 0.016$ \cite{accms11}. The
uncertainty comprises all sources of uncertainties.
Fig.\ref{ac} shows the dependence of $A_{C}$ on CMDM $(\kappa)$. The
dashed line is the upper limit on $A_{C}$ measured by the CMS collaboration
 at the LHC. This provides very loose bounds on $\kappa$. Charge
 asymmetry is almost insensitive to $\tilde{\kappa}$ and therefore
 does not give notable limits on $\tilde{\kappa}$.

Finally, in Table \ref{results1} we compare the limits obtained from
the LHC and Tevatron cross sections, the ratio of gluon-gluon fusion
in top pair production rate ($R$), and the charge asymmetry. The
bounds from ratio $R$ are not strong but precise
measurement of the ratio $R$ could provide strong limits on
$\kappa$ and $\tilde{\kappa}$.
Charge asymmetry do not show large sensitivity to CEDM and CMDM.

\begin{table}
\begin{center}
\begin{tabular}{|c|c|c|c|c|}
  \hline
           & $\sigma_{LHC}$ and $d\sigma_{TeV}/dm_{t\bar{t}}$ & ratio
           $R$ & $A_{C}$ \\ \hline
 $\kappa $ & $-0.03<\kappa<0.01$                              & $-1.1<\kappa<0.6$ & $-1.72<\kappa<2.02$ \\ \hline
 $\tilde{\kappa}$ & $-0.063<\tilde{\kappa}<0.063$&
                  $-0.8<\tilde{\kappa}<0.8$& - \\
  \hline
\end{tabular}
\caption{\label{results1} Comparison of the bounds extracted from
  Tevatron, LHC, gluon-gluon fusion ratio, and $A_{C}$ on $\kappa$ and
  $\tilde{\kappa}$.}
\end{center}
\end{table}

\section{Conclusions}

In this paper, first we obtained bounds on the chromoelectric and
chromomagnetic dipole moments of the top quark using the present
measured total
cross section of $t\bar{t}$ at the LHC and the diferential cross
setion of $t\bar{t}$ in invariant mass of the top pair system
($d\sigma/dm_{t\bar{t}}$) at the Tevatron. A global $\chi^{2}$ fit over the invariant
mass bins from Tevatron and the measured total cross sections at the LHC has
been performed.
Because of precise measurements of the total and differential cross
sections, the bounds obtained from the $\chi^{2}$ are stringent.
Then we studied the effects of anomalous
top quark coupling with gluon on the fraction of gluon-gluon fusion in
the $t\bar{t}$ production cross section at the Tevatron and LHC.
We found that the fraction of gluon-gluon fusion in $t\bar{t}$
production rate $(R)$ at the Tevatron is more sensitive to CEDM and
CMDM than the LHC. It is shown that the precise enough measurement of
$R$ could provide bounds on CMDM ($\kappa$) and CEDM($\tilde{\kappa}$)
comparable with the bounds from cross sections. We also examined the
sensitivity of charge asymmetry ($A_{C}$) in $t\bar{t}$ events at the LHC to
$\kappa$ and $\tilde{\kappa}$. We found that the presence of
$\tilde{\kappa}$ does not produce any charge asymmetry but the current limit on $A_{C}$ gives
loose bounds on $\kappa$.

%\vspace{0.1cm}
{\bf Acknowledgments}\\
We are grateful to S. Khatibi for useful discussions.


\begin{thebibliography}{99}
\bibitem{wagner}
  W.~Wagner,
  %``Top quark physics in hadron collisions,''
  Rept.\ Prog.\ Phys.\  {\bf 68}, 2409 (2005)
  [hep-ph/0507207].
  %%CITATION = HEP-PH/0507207;%%
\bibitem{werner} W.~ Bernreuther, arXiv:0805.1333 [hep-ph].
\bibitem{beneke} M.~Beneke {\it et al.}, arXiv:hep-ph/0003033.


\bibitem{higgs1} R.~Martinez and J.~A.~Rodriguez,
  %``The Anomalous chromomagnetic dipole moment of the top quark in the standard model and beyond,''
  Phys.\ Rev.\ D {\bf 65}, 057301 (2002)
  [hep-ph/0109109].
  %%CITATION = HEP-PH/0109109;%%



\bibitem{higgs2} R.~D.~Peccei and X.~Zhang,
  %``Dynamical Symmetry Breaking and Universality Breakdown,''
  Nucl.\ Phys.\ B {\bf 337}, 269 (1990);   R.~D.~Peccei, S.~Peris and X.~Zhang,
  %``Nonstandard couplings of the top quark and precision measurements of the electroweak theory,''
  Nucl.\ Phys.\ B {\bf 349}, 305 (1991).
  %%CITATION = NUPHA,B349,305;%%


\bibitem{k1}
  K.~-m.~Cheung,
  %``Probing the chromoelectric and chromomagnetic dipole moments of the top quark at hadronic colliders,''
  Phys.\ Rev.\ D {\bf 53}, 3604 (1996)
  [hep-ph/9511260]; K. Cheung,  Phys.\ Rev.\ D {\bf 55}, 4430 (1997).
  %%CITATION = HEP-PH/9511260;%%


\bibitem{k2}
 P.~Haberl, O.~Nachtmann and A.~Wilch,
  %``Top production in hadron hadron collisions and anomalous top - gluon couplings,''
  Phys.\ Rev.\ D {\bf 53}, 4875 (1996)
  [hep-ph/9505409].
  %%CITATION = HEP-PH/9505409;%%

\bibitem{k3}
  T.~G.~Rizzo,
  %``Single top quark production as a probe for anomalous moments at hadron colliders,''
  Phys.\ Rev.\ D {\bf 53}, 6218 (1996)
  [hep-ph/9506351].
  %%CITATION = HEP-PH/9506351;%%

\bibitem{k4}
 D.~Atwood, A.~Kagan and T.~G.~Rizzo,
  %``Constraining anomalous top quark couplings at the Tevatron,''
  Phys.\ Rev.\ D {\bf 52}, 6264 (1995)
  [hep-ph/9407408].
  %%CITATION = HEP-PH/9407408;%%

\bibitem{k5}
  T.~G.~Rizzo,
  %``Top quark production at the Tevatron: Probing anomalous chromomagnetic moments and theories of low scale gravity,''
  hep-ph/9902273.
  %%CITATION = HEP-PH/9902273;%%

\bibitem{k6}
  K.~-i.~Hikasa, K.~Whisnant, J.~M.~Yang and B.~-L.~Young,
  %``Probing anomalous top quark interactions at the Fermilab Tevatron collider,''
  Phys.\ Rev.\ D {\bf 58}, 114003 (1998)
  [hep-ph/9806401].
  %%CITATION = HEP-PH/9806401;%%

\bibitem{k61}
  D.~Choudhury and P.~Saha,
  %``Probing Top Anomalous Couplings at the Tevatron and the Large Hadron Collider,''
  Pramana {\bf 77}, 1079 (2011)
  [arXiv:0911.5016 [hep-ph]].
  %%CITATION = ARXIV:0911.5016;%%



\bibitem{k7}

 Z.~HIOKI and K.~OHKUMA,
  %``Exploring anomalous top interactions via the final lepton in $t\bar{t}$ productions/decays at hadron colliders,''
  Phys.\ Rev.\ D {\bf 83}, 114045 (2011)
  [arXiv:1104.1221 [hep-ph]]; Z.~Hioki and K.~Ohkuma,
  %``Optimal-observable Analysis of Possible Non-standard Top-quark Couplings in pp -> t t-bar X -> l^+ X',''
  arXiv:1206.2413 [hep-ph].

\bibitem{dmtt}
  T.~Aaltonen {\it et al.}  [CDF Collaboration],
  %``First Measurement of the t anti-t Differential Cross Section d sigma/dM(t anti-t) in p anti-p Collisions at s**(1/2)=1.96-TeV,''
  Phys.\ Rev.\ Lett.\  {\bf 102}, 222003 (2009)
  [arXiv:0903.2850 [hep-ex]].
  %%CITATION = ARXIV:0903.2850;%%




\bibitem{NLC}

  T.~G.~Rizzo,
  %``Probing anomalous chromomagnetic top quark couplings at the NLC,''
  Phys.\ Rev.\ D {\bf 50}, 4478 (1994)
  [hep-ph/9405391].
  %%CITATION = HEP-PH/9405391;%%


\bibitem{bsg1}

  R.~Martinez and J.~A.~Rodriguez,
  %``Using the radiative decay $b \to s \gamma$ to bound the chromomagnetic dipole moment of the top quark,''
  Phys.\ Rev.\ D {\bf 55}, 3212 (1997)
  [hep-ph/9612438].
  %%CITATION = HEP-PH/9612438;%%


\bibitem{bsg2}

  R.~Martinez and J.~A.~Rodriguez,
  %``The Anomalous chromomagnetic dipole moment of the top quark in the standard model and beyond,''
  Phys.\ Rev.\ D {\bf 65}, 057301 (2002)
  [hep-ph/0109109].
  %%CITATION = HEP-PH/0109109;%%


\bibitem{mssm}
  J.~-M.~Yang and C.~-S.~Li,
  %``Supersymmetric electroweak corrections to top quark production at the Fermilab Tevatron,''
  Phys.\ Rev.\ D {\bf 52}, 1541 (1995)
  [Erratum-ibid.\ D {\bf 54}, 3671 (1996)];   J.~M.~Yang and C.~S.~Li,
  %``Top squark mixing effects in the supersymmetric electroweak corrections to top quark production at the Tevatron,''
  Phys.\ Rev.\ D {\bf 54}, 4380 (1996)
  [hep-ph/9603442].


\bibitem{2hdm}
  C.~S.~Li, R.~J.~Oakes and J.~M.~Yang,
  %``Yukawa corrections to single top quark production at the Fermilab Tevatron in the two Higgs doublet models,''
  Phys.\ Rev.\ D {\bf 55}, 1672 (1997)
  [hep-ph/9608460].
  %%CITATION = HEP-PH/9608460;%%

\bibitem{LH}
 Q.~-H.~Cao, C.~-R.~Chen, F.~Larios and C.~-P.~Yuan,
  %``Anomalous gtt couplings in the Littlest Higgs Model with T-parity,''
  Phys.\ Rev.\ D {\bf 79}, 015004 (2009)
  [arXiv:0801.2998 [hep-ph]]; L.~Ding and C.~-X.~Yue,
  %``Top quark chromomagnetic dipole moment in the littlest Higgs model with T-parity,''
  Commun.\ Theor.\ Phys.\  {\bf 50}, 441 (2008)
  [arXiv:0801.1880 [hep-ph]].

\bibitem{unparticle}
  R.~Martinez, M.~A.~Perez and O.~A.~Sampayo,
  %``Constraints on unparticle physics from the gt $\bar{t}$ anomalous coupling,''
  Int.\ J.\ Mod.\ Phys.\ A {\bf 25}, 1061 (2010)
  [arXiv:0805.0371 [hep-ph]].
  %%CITATION = ARXIV:0805.0371;%%




\bibitem{cteq}
  J.~Pumplin, D.~R.~Stump, J.~Huston, H.~L.~Lai, P.~M.~Nadolsky and W.~K.~Tung,
  %``New generation of parton distributions with uncertainties from global QCD
  %analysis,''
  JHEP {\bf 0207}, 012 (2002)
  [arXiv:hep-ph/0201195].
  %%CITATION = JHEPA,0207,012;%%


\bibitem{lhc} The CMS Collaboration, CMS PAS TOP-11-005.


\bibitem{tevatron} The CDF Collaboration, Conf. Note 9913.

\bibitem{sigmatheory}
  N.~Kidonakis,
  %``Differential and total cross sections for top pair and single top production,''
  arXiv:1105.3481 [hep-ph].
  %%CITATION = ARXIV:1205.3453;%%

\bibitem{lhc8} The CMS Collaboration, CMS PAS TOP-12-006.


\bibitem{cacciari}
 M.~Cacciari, S.~Frixione, M.~L.~Mangano, P.~Nason and G.~Ridolfi,
  %``The t anti-t cross-section at 1.8-TeV and 1.96-TeV: A Study of the systematics due to parton densities and scale dependence,''
  JHEP {\bf 0404}, 068 (2004)
  [hep-ph/0303085].
  %%CITATION = HEP-PH/0303085;%%

\bibitem{ggbsm1}  C.~T.~Hill and S.~J.~Parke,
  %``Top production: Sensitivity to new physics,''
  Phys.\ Rev.\ D {\bf 49}, 4454 (1994)
  [hep-ph/9312324];   K.~D.~Lane and E.~Eichten,
  %``Natural topcolor assisted technicolor,''
  Phys.\ Lett.\ B {\bf 352}, 382 (1995)
  [hep-ph/9503433].



\bibitem{gg}
  T.~Aaltonen {\it et al.}  [CDF Collaboration],
  %``Measurement of the fraction of $t \bar{t}$ production via gluon-gluon fusion in $p \bar{p}$ collisions at $\sqrt{s}$ = 1.96-TeV,''
  Phys.\ Rev.\ D {\bf 79}, 031101 (2009)
  [arXiv:0807.4262 [hep-ex]].
  %%CITATION = ARXIV:0807.4262;%%


\bibitem{AC11}
  J.~H.~Kuhn and G.~Rodrigo,
  %``Charge asymmetries of top quarks at hadron colliders revisited,''
  JHEP {\bf 1201}, 063 (2012)
  [arXiv:1109.6830 [hep-ph]].
  %%CITATION = ARXIV:1109.6830;%%


\bibitem{accms11} The CMS Collaboration, CMS-PAS-TOP-11-030.


\end{thebibliography}
\end{document}